\documentclass[aps,prb,preprint,showpacs,showkeys,floatfix]{revtex4}

\usepackage{graphicx,color}
\usepackage{natbib}
\usepackage[utf8]{inputenc}
\usepackage{float}
\graphicspath{{figs/}}
\usepackage{amsmath}
\usepackage{tabularx}
\usepackage[colorlinks,linkcolor=blue,anchorcolor=blue,citecolor=blue]{hyperref}
\begin{document}

\title{Tunable mechanical properties of the graphene/MoS$_2$ tubal van der Waals heterostructure}

\author{Ya-Wen Tan}
    \affiliation{Shanghai Institute of Applied Mathematics and Mechanics, Shanghai Key Laboratory of Mechanics in Energy Engineering, Shanghai University,~Shanghai~200072,~People's~Republic~of~China}

\author{Jin-Wu Jiang}
    \altaffiliation{Corresponding author: jwjiang5918@hotmail.com}
    \affiliation{Shanghai Institute of Applied Mathematics and Mechanics, Shanghai Key Laboratory of Mechanics in Energy Engineering, Shanghai University,~Shanghai~200072,~People's~Republic~of~China}

\date{\today}
\begin{abstract}

We propose a tubal van der Waals heterostructure by rolling up the graphene and MoS$_2$ atomic layers into a tubal form. We illustrate that the interlayer space for the tubal van der Waals heterostructure can be varied in a specific range, which is determined by the competition between the interlayer van der Waals force and the mechanical properties of the atomic layers. The variability of the interlayer space can be utilized to efficiently tune mechanical properties of the tubal van der Waals heterostructure. More specifically, we demonstrate that the Poisson$^{\prime}$s ratio of the tubal van der Waals heterostructure can be manipulated by a factor of two by varying the interlayer space from 1.44~{\AA} to 4.44~{\AA}. Our work promotes a new member with tunable Poisson$^{\prime}$s ratio to the van der Waals heterostructure family.

\end{abstract}

\maketitle
\pagebreak

\section{Introduction}

Two-dimensional atomic and molecular membranes represent the thinnest materials in nature.\cite{GeimAK2007nm} This intrinsic thinness enables them to bend and fold easily, as the bending rigidity scales with thickness cubed.\cite{LandauLD}  When combined with the fact that most two-dimensional materials have a high in-plane stiffness, the out of plane flexibility of these two-dimensional atomic membranes enables them to conform to rough surfaces, and exhibit two-dimensional to three-dimensional shape transitions with interesting implications for flexible and stretchable electronics.\cite{KhangDY2006sci,KimDH2008sci,RogersJA2010sci}

The vertical van der Waals heterostructures have been of interest for some time due to the notion that stacking two-dimensional materials could lead to novel physical properties and potential applications.\cite{GeimAK2013nat,DengD2016nn,NovoselovKS2016sci} For instance, the graphene/MoS$_2$ vertical van der Waals heterostructure can be used as hybrid photoactive device, in which the MoS$_2$ has enhanced photon absorption and electron-hole creation while graphene serves as a transport electrode.\cite{BritnellL2013sci} In other words, the advanced properties of individual atomic layers can be integrated into the van der Waals heterostructure by vertically stacking these atomic layers.

As another important low-dimensional material, the core-shell nanowire is one fundamental building block for one-dimensional nano-materials.\cite{ZhangA2016} The core-shell nanowire also combines the advantages of the core and shell nanowires. For example, the crystalline-amorphous core-shell silicon nanowire\cite{CuiLF2008nl} and the carbon-silicon core-shell nanowire\cite{CuiLF2009nl} are both promising candidates for electrode materials of high capacity. The metal oxide core-shell nanowire arrays were proposed for electrochemical energy storage.\cite{XiaX2012acsnn}

Basically, these novel performance of the core-shell nanowire are in close relation to its heterostructure configuration in the radial direction; i.e., the different functionalities of the core and shell nanowires are integrated to exhibit novel properties. As inspired by the one-dimensional core-shell nanowire structure, we propose a new type of van der Waals heterostructure in the tubal form for the two-dimensional atomic layers.

In this paper, we propose a tubal van der Waals heterostructure for the graphene and MoS$_2$ atomic layers. This tubal heterostructure has an additional degree of freedom as compared with the vertical van der Waals heterostructure; i.e., the interlayer space is variable in the range of [1.44, 4.44]~{\AA}. These lower and upper boundaries for the interlayer space are determined by the specific mechanical properties of graphene and MoS$_2$. The variation of the interlayer space can be used to manipulate mechanical properties for the tubal van der Waals heterostructure. We illustrate the tunability of the Poisson$^{\prime}$s ratio by a factor of two with the interlayer space varying in the range [1.44, 4.44]~{\AA}.

\section{Structure and Simulation Details}

As we have discussed in the introduction section, graphene and MoS$_2$ are two most promising nano-materials that exhibit lots of novel physical and mechanical properties. Graphene is metalic without intrinsic electronic band gap, but it has many wonderful mechanical properties, including an ultra-high in-plane mechanical stiffness and extremely small bending modulus. On the other hand, MoS$_2$ is a good semiconductor with intrinsic finite electronic band gap. As a result, graphene, MoS$_2$ and other two-dimensional materials have been stacked into functional vertical van der Waals heterostructures, in which graphene serves as a protection layer while MoS$_2$ as electronic or optical layer.

In the vertical van der Waals heterostructure, the interlayer space is a constant that is determined by the neighboring atomic layers. We herein propose a tubal heterostructure consistuted by graphene and MoS$_2$ as shown in Fig.~\ref{fig_def}~(a). The inner layer is a MoS$_2$ tube while the outer layer is a carbon nanotube (CNT). There shall be several advanced features for this tubal heterostructure. First of all, the outer CNT can protect the inner MoS$_2$ tube from external influence. Furthermore, the tubal structure adds a constraint in the radial direction for the MoS$_2$, so this type of heterostructure shall be stable. In the present work, the inner MoS$_2$ tube and the outer CNT are both along the armchair orientation. The length of the graphene/MoS$_2$ (GM) tubal heterostructure along the axis is 21.7~{\AA}, and we will show that the length of the GM tubal heterostructure does not affect its mechanical properties with periodic boundary condition applied in the axial direction.

The interaction among carbon atoms is described by the second generation Brenner potential~\cite{brennerJPCM2002}. The interaction within the MoS$_2$ is described by the Stillinger-Weber potential.\cite{JiangJW2017intech} The van der Waals interaction between the inner MoS$_2$ tube and the outer CNT is captured by the following Lennar-Jones potential,
\begin{equation}
V_{LJ}=4\epsilon\left[\left(\frac{\sigma}{r}\right)^{12}-(\frac{\sigma}{r})^{6}\right],
\label{eq_lj}
\end{equation}
where $\epsilon=0.00836$~{eV} and $\sigma=3.28$~{\AA} are the parameters.\cite{JiangJW2015jap} The distance between two atoms is $r$. The periodic boundary condition is applied to the axial direction. The structure is thermalized to the thermal steady state with the NPT (constant particle number, constant pressure, and constant temperature) ensemble for 100~ps by the Nos\'e-Hoover approach.\cite{Nose,Hoover} In the simulation of the tensile process, the tubal heterostructure is stretched along the axial direction at a strain rate of $10^7$~{s$^{-1}$}, while the stress in the radial direction is allowed to be relaxed to be zero. The present simulations are performed at 4.2~K, expect for the investigation of the temperature effect. The standard Newton equations of motion are integrated in time using the velocity Verlet algorithm with a time step of 1~{fs}. Simulations are performed using the publicly available simulation code LAMMPS~\cite{PlimptonSJ}. The OVITO package is used for visualization~\cite{ovito}.

\section{Interlayer space of the tubal heterostructure}

\begin{figure}[H]
 \centering
    \scalebox{1.1}[1.1]{\includegraphics[width=8cm]{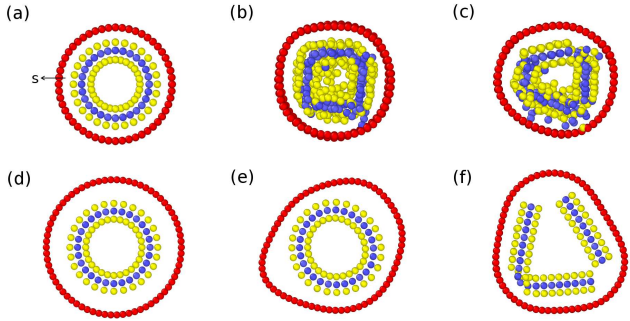}}
  \caption{Two typical collapse styles for the GM tubal heterostructure. Style I: (a), (b), and (c) are the initial, intermediate, and final relaxed configurations for the GM tubal heteorsturcture with small interlayer space ($s$) of 1.44~{\AA}. Style II: (d), (e), and (f) are the initial, intermediate, and final relaxed configurations for the GM tubal heterostructure with large interlayer space of 5.44~{\AA}. The length of the GM tubal heterostructure is 21.7~{\AA}. The inner MoS$_2$ has the diameter of 14~{\AA}. Atoms C, Mo, and S are represented by red, blue, and yellow colors.}
  \label{fig_def}
\end{figure}

Different from the constant interlayer space in the vertical van der Waals heterostructure, there is typically a variation range for the value of the interlayer space in the tubal structure. For example, the interlayer space for the double-walled CNT can vary between [3.4, 4.1]~{\AA} in Ref~\onlinecite{RenW2002cpl} or [3.4, 4.7]~{\AA} in Ref~\onlinecite{LiL2005carbon}. Similar interlayer space was also found for the double-walled CNT in other experiments.\cite{ZhuH2002carbon,CiL2002cpl} We thus first examine the variation range for the interlayer space of the GM tubal heterostructure.

For a given inner MoS$_2$ tube, the outer layer can be constructed by the CNT of different diameter, which results in different interlayer space. We investigate the lower and upper boundaries for the interlayer space. Obviously, there should be a lower boundary for the interlayer space, as shown in Fig.~\ref{fig_def}. In Fig.~\ref{fig_def}~(a), the diameter of the inner MoS$_2$ tube is 14~{\AA}, and the initial value for the interlayer space is 1.44~{\AA}. We perform energy minimization for the initial configuration of the GM tubal heterostructure. Fig.~\ref{fig_def}~(b) and (c) show that the inner MoS$_2$ tube is cracked by the large repulsive van der Waals force as a result of the small space between the inner MoS$_2$ tube and the outer CNT. Note that the outer CNT is not damaged owing to its super high in-plane stiffness. For the inner MoS$_2$ tube of diameter 14~{\AA}, we find that the lower boundary of the interlayer space is 1.44~{\AA}, bellow which the GM tubal heterostructure becomes unstable as the inner MoS$_2$ tube will be cracked by the strong interlayer force.

We also find that there is an upper boundary for the interlayer space of the GM tubal heterostructure. In Fig.~\ref{fig_def}~(d), the diameter of the inner MoS$_2$ tube is 14~{\AA}, while the initial interlayer space is 5.44~{\AA}. Fig.~\ref{fig_def}~(e) and (f) illustrate two configurations during the energy minimization process. Panel (e) shows that the shape of the outer CNT is distorted by the attractive interlayer van der Waals force. The interlayer force is weak, as the interlayer space between the inner MoS$_2$ tube and outer CNT is large. However, this weak force is still able to destroy the circular shape of the cross-section for the outer CNT layer, because of the ultra small bending modulus of the carbon sheet. The deformation of the outer CNT layer can generate the compressive force on the inner MoS$_2$ tube, and eventually crack the brittle MoS$_2$ tube as shown in Fig.~\ref{fig_def}~(f). For the inner MoS$_2$ tube of diameter 14~{\AA}, we find that the upper boundary for the interlayer space of the GM tubal heterostructure is 3.9~{\AA}, above which the heterostructure will be unstable.

\begin{figure}[H]
 \centering
    \scalebox{1.1}[1.1]{\includegraphics[width=8cm]{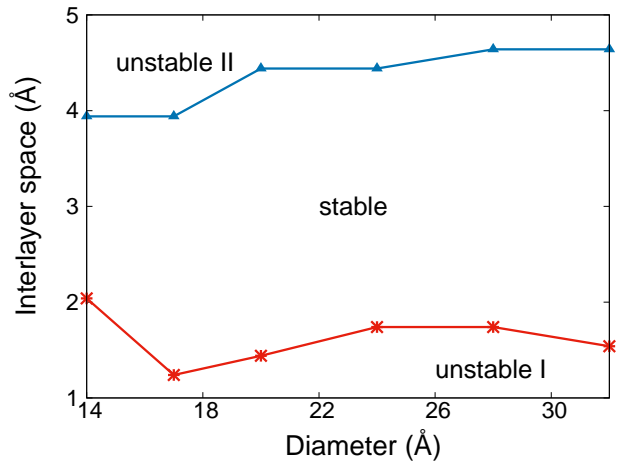}}
  \caption{The lower and upper boundaries for the interlayer space of the GM tubal heterostructure with the inner MoS$_2$ tube of different diameter.}
  \label{fig_stab}
\end{figure}

\begin{figure}[H]
 \centering
    \scalebox{1.1}[1.1]{\includegraphics[width=8cm]{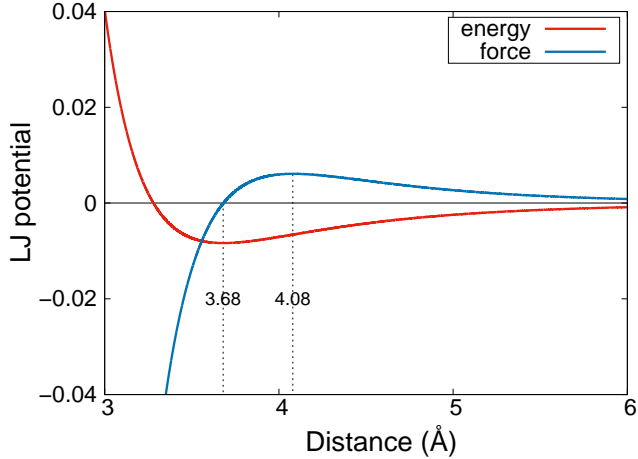}}
  \caption{The distance dependence for the Lennard-Jones potential and force. Note that the force is zero at 3.68~{\AA}, while the maximum force locates at 4.08~{\AA}.}
  \label{fig_lj}
\end{figure}

We show in Fig.~\ref{fig_stab} the lower and upper boundaries for the interlayer space in the GM tubal heterostructure with the inner MoS$_2$ tube of different diameter. The lower boundary of the interlayer space fluctuates around 1.44~{\AA} while the upper boundary is around 4.44 ~{\AA} for the GM tubal heterostructure of large diameter, which is related to the van der Waals force between MoS$_2$ and CNT. The van der Waals force ($V_{lj}^{\prime}$) is plotted in Fig.~\ref{fig_lj} as a function of the interatomic distance. The van der Waals force achieves the maximum value at the distance of 4.08~{\AA}, which is  close to the upper boundary for the interlayer space. It means that the GM tubal heterostructures with interlayer space of 4.44~{\AA} will have the maximum interlayer van der Waals force, so these structures are unstable. It should be noted that the van der Waals force is weaker for interlayer space larger than the upper boundary value. However, the outer CNT has larger diameter for large interlayer space, so it is easier to be distorted even though the interlayer van der Waals force is weaker. As a result, the GM tubal heterostructure is unstable if the interlayer space is larger than the upper boundary value. The GM tubal heterostructures with interlayer space value in the region between the two curves in Fig.~\ref{fig_stab} are stable and will be considered for further simulations in the rest of the manuscript.

\begin{figure}[H]
 \centering
    \scalebox{1.1}[1.1]{\includegraphics[width=8cm]{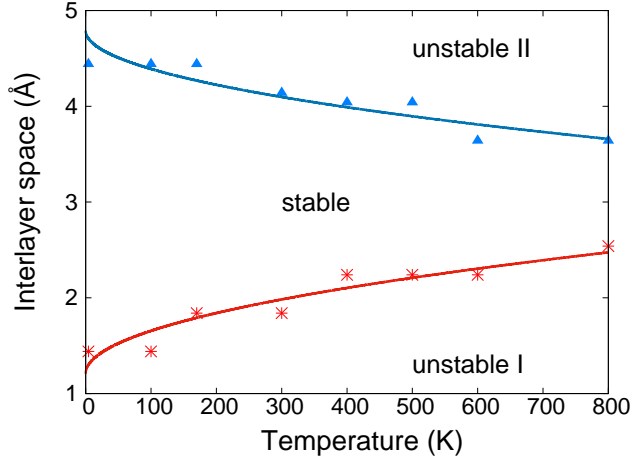}}
  \caption{The temperature dependence for the lower and upper boundaries of the GM tubal heterostructure. The diameter of the inner MoS$_2$ tube is 20.0~{\AA}. Note that the stable region of the interlayer space is narrowed with increasing temperature. Numerical results are fitted to the function $s=a+b\sqrt{T}$. See text for the physical origin of this fitting function.}
  \label{fig_stab1}
\end{figure}

We now explore the temperature effect on the lower and upper boundaries of the interlayer space for the GM tubal heterostructure. The inner MoS$_2$ tube of the GM tubal heterostructure has a diameter of 20.0~{\AA} in this set of simulations. The lower and upper boundaries at different temperatures are shown in Fig.~\ref{fig_stab1}. The stable region (which is sandwiched by the lower and upper boundaries) for the interlayer space is narrowed with the increase of the temperature. It is because with increasing temperature, the thermal vibration becomes stronger thus leading to the increase of the possibility for the collapse of the structure. More specifically, the thermal-induced vibrational amplitude ($A$) is proportional to the square root of temperature $T$; i.e., $A\propto\sqrt{T}$. The possibility for the collapse of the structure is proportional to the vibrational amplitude $A$. Hence, the temperature dependence for the lower and upper boundaries shall be fitted to the function $s=a+b\sqrt{T}$, with $a$ and $b$ as fitting parameters.

Indeed, Fig.~\ref{fig_stab1} verifies that the lower boundary can be well fitted to $s_{\rm lower}=1.2+0.04\sqrt{T}$, while the upper boundary is fitted to $s_{\rm upper}=4.8-0.04\sqrt{T}$. At room temperature, the lower boundary is 1.9~{\AA} from our theoretical result, which is smaller than the experimental value for the lower boundary of the interlayer space of the double-walled CNT.\cite{RenW2002cpl,LiL2005carbon} It is probably because there are some defects or impurities in the experimental samples, which have negative effects on the stability of the structure. The upper boundary at the room temperature is 4.2~{\AA}, which is very close to the experimental value for the double-walled CNT.\cite{RenW2002cpl,LiL2005carbon}

\section{Tunable Poisson$^{\prime}$s ratio via interlayer space}

We have shown in above that the interlayer space in the GM tubal heterostructure is not necessarily a constant. Instead, the interlayer space can be variable in the range of [1.44, 4.44]~{\AA} at 4.2~K. Taking advantage of the variation of the interlayer space, we may be able to tune some physical or mechanical properties for the GM tubal heterostructure. As an example, we investigate the Poisson$^{\prime}$s ratio of the GM tubal heterostructure with different interlayer space. The Poisson$^{\prime}$s ratio for the tubal heterostructure is computed by
\begin{equation}
\nu=-\frac{\epsilon_r}{\epsilon_x},
\label{eq_poisson}
\end{equation}
where $\epsilon_x \in [0, 0.02]$ is the tensile strain applied along the axial direction. The strain in the radial direction is calculated by measuring the variation of the radius $r$ of outer CNT layer in the tubal heterostructure; i.e., $\epsilon_r=\Delta r/r_0$.

\begin{figure}[H]
 \centering
    \scalebox{1.1}[1.1]{\includegraphics[width=8cm]{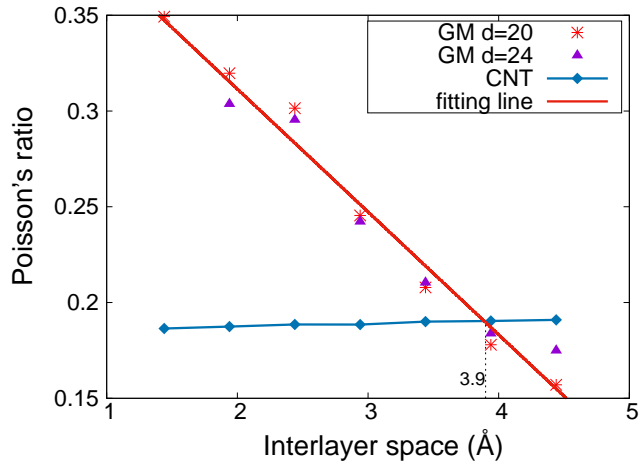}}
  \caption{The dependence of the Poisson$^{\prime}$s ratio ($\nu$) on the interlayer space ($s$) for the GM tubal heterostructure. The Poisson$^{\prime}$s ratio is well fitted to the linear function $\upsilon=0.44-0.06s$.}
  \label{fig_pos}
\end{figure}

The effect of the interlayer space on the Poisson$^{\prime}$s ratio of the GM tubal heterostructure is illustrated in Fig.~\ref{fig_pos}. The diameter of the inner MoS$_2$ tube is 20.0~{\AA} or 24.0~{\AA} in this set of calculations. The Poisson$^{\prime}$s ratio for the single-walled CNT is also shown in the figure for comparison. For a given interlayer space, the corresponding single-walled CNT is determined by setting the same diameter as the outer CNT layer in the GM tubal heterostructure. The Poisson$^{\prime}$s ratio for the single-walled CNT is almost a constant, as the diameter of these CNTs is larger. The Poisson$^{\prime}$s ratio for the GM tubal heterostructrue is not sensitive to the diameter of the inner MoS$_2$ tube. The Poisson$^{\prime}$s ratio can be well fitted to a single linear function $\nu=0.44-0.06s$ for these two sets of GM tubal heterostructures with the inner MoS$_2$ tube diameter as 20.0~{\AA} or 24.0~{\AA}.

The linear relation between the Poisson$^{\prime}$s ratio and the interlayer space also indicates that the Poisson$^{\prime}$s ratio of the GM tubal heterostructure is tunable in a wide range via the interlayer space. More specifically, the Poisson$^{\prime}$s ratio decreases with increasing interlayer space, and the Poisson$^{\prime}$s ratio for the tubal heterostructure with interlayer space of 1.44~{\AA} is about twice of that for the interlayer space of 4.44~{\AA}. At the interlayer space around $s=3.9$~{\AA}, the GM tubal heterostructure has the same Poisson$^{\prime}$s ratio value as the single-walled CNT. For $s>3.9$~{\AA}, the Poisson$^{\prime}$s ratio of the GM tubal heterostructure decreases with increasing interlayer space. For $s<3.9$~{\AA}, the Poisson$^{\prime}$s ratio of the heterostructure increases with decreasing of the interlayer space. We note that, in the vertically stacking GM heterostructure, the interlayer space is a constant 3.2~{\AA}, and the Poisson$^{\prime}$s ratio is 0.29. As a result, the tunable interlayer space in the GM tubal heterostructure is an advantage over the GM van der Waals vertical heterostructure.

The difference between the Poisson$^{\prime}$s ratio of the GM tubal heterostructure and the single-walled CNT is caused by the van der Waals interaction between the inner MoS$_2$ tube and the outer CNT layer in the GM tubal heterostructure. Fig.~\ref{fig_pos} shows that the GM tubal heterostructure with interlayer space $s=3.9$~{\AA} has the same Poisson$^{\prime}$s ratio as the CNT, because the van der Waals force is zero at the distance of 3.68~{\AA} (which is close to the intersecting point of 3.9~{\AA}) as shown in Fig.~\ref{fig_lj}.

For the GM tubal heterostructure with interlayer space $s>3.9$~{\AA}, the van der Waals force is attractive for the outer CNT layer, which is balanced by the in-plane compressive force within the outer CNT layer. The compressive force is caused by the compression of the outer CNT layer along the circumference direction, so it is determined by the in-plane Young's modulus (which is large) of the outer CNT layer. It is thus reasonable to assume that the variation of the in-plane compressive force will be more important than the variation of the van der Waals force during the axial streching of the GM tubal heterostructure. The in-plane compressive force has the effect to expand the outer CNT layer, which counteracts with the Poisson effect induced contraction of the outer CNT layer. As a result, the Poisson$^{\prime}$s ratio for the GM tubal heterostructure with interlayer space $s>3.9$~{\AA} is smaller than the single-walled CNT. The in-plane compressive force is larger for the GM tubal heterostructure with larger interlayer space for $s>3.9$~{\AA}, which results in the decrease of the Poisson$^{\prime}$s ratio for the heterostructure with further increasing interlayer space.

Analogously, for the GM tubal heterostructure with interlayer space $s<3.9$~{\AA}, the in-plane force within the outer CNT layer is tensile, so it will assist the Poisson effect induced contraction of the outer CNT layer. As a result, the heterostructure has larger Poisson$^{\prime}$s ratio than the single-walled CNT for $s<3.9$~{\AA}. The in-plane tensile force becomes larger with further decreasing interlayer space for $s<3.9$~{\AA}, which leads to the increase of the Poisson$^{\prime}$s ratio for the heterostructure with decreasing interlayer space.

\begin{figure}[H]
 \centering
    \scalebox{1.1}[1.1]{\includegraphics[width=8cm]{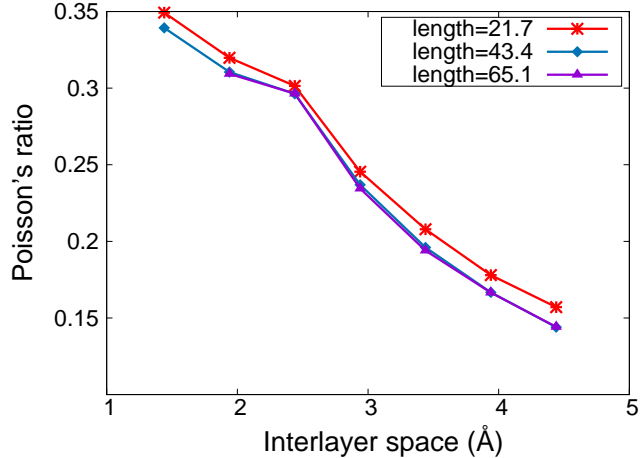}}
  \caption{The dependence of the Poisson$^{\prime}$s ratio on the interlayer space for the GM tubal heterostructures of defferent length.}
  \label{fig_length}
\end{figure}

We note that, in the above, the length of the GM heterostructure is set to be 21.7~{\AA}, while periodic boundary condition is used along the axial direction. We show in Fig.~\ref{fig_length} that the length has almost no effect on the Poisson$^{\prime}$s ratio.

\section{Conclusion}
To summary, we propose a new type of van der Waals heterostructure based on graphene and MoS$_2$, in which the atomic layers are stacked in the tubal form. A distinct feature of the tubal heterostructure is its variable interlayer space, which shall be constant in the heterostructure of the usual vertical form. However, we demonstrate that the interlayer space is within the range of [1.44, 4.44]~{\AA}, which is governed by the interplay among the mechanical properties of the MoS$_2$ and graphene, and the interlayer van der Waals interaction. The variability of the interlayer space can be used to tune physical or mechanical properties for the tubal heterostructure. As an example, we illustrate that the Poisson$^{\prime}$s ratio of the GM tubal heterostructure can be tuned by a factor of two by varying the interlayer space in the range of [1.44, 4.44]~{\AA}. The inter-layer space shall have similar important effect on other physical or mechanical properties, like the electronic conductance and thermal transport of the tubal heterostructure.

\textbf{Acknowledgment} The work is supported by the Recruitment Program of Global Youth Experts of China, the National Natural Science Foundation of China (NSFC) under Grant No. 11504225, and the Innovation Program of Shanghai Municipal Education Commission under Grant No. 2017-01-07-00-09-E00019.


\end{document}